\documentclass[reprint,twocolumn,superscriptaddress,showpacs,showkeys,amsmath,amssymb,floatfix,aps,prl,nolongbibliography]{revtex4-1}
\usepackage[english]{babel}
\usepackage[utf8]{inputenc}
\usepackage{graphicx}
\usepackage{fancyhdr}
\usepackage[active]{srcltx}
\usepackage{setspace}
\usepackage{float}
\usepackage{color}
\usepackage{multirow}
\usepackage{bm}
\usepackage{amsmath, amsfonts, amssymb, mathtools}	
\usepackage{latexsym}
\usepackage{soul}
\usepackage{gensymb}

\usepackage[breaklinks, hidelinks]{hyperref}
\usepackage{xcolor}
\hypersetup{
    colorlinks,
    linkcolor={blue!70!black},
    citecolor={blue!70!black},
    urlcolor={blue!70!black}
}

\newcommand{\LI}{\lambda_{I}}
\newcommand{\LR}{\lambda_{R}}

\renewcommand\vec[1]{\textit{\textbf{#1}}}

\begin{document}

\title{Emergence of radial Rashba spin-orbit fields in twisted van der Waals heterostructures}

\author{Tobias \surname{Frank}}

\author{Paulo E. \surname{Faria~Junior}}
\email[Email to: ]{fariajunior.pe@gmail.com}

\author{Klaus \surname{Zollner}}

\author{Jaroslav \surname{Fabian}}

\affiliation{Institute for Theoretical Physics, University of Regensburg, 93040 Regensburg, Germany}

\date{\today}

\begin{abstract}
Rashba spin-orbit coupling is a quintessential spin interaction appearing in virtually any electronic heterostructure. Its paradigmatic spin texture in the momentum space forms a tangential vector field. Using first-principles investigations, we demonstrate that in twisted homobilayers and hetero-multilayers, the Rashba coupling can be predominantly radial, parallel to the momentum. Specifically, we study four experimentally relevant structures: twisted bilayer graphene (Gr), twisted bilayer WSe$_2$, and twisted multilayers WSe$_2$/Gr/WSe$_2$ and WSe$_2$/Gr/Gr/WSe$_2$. We show, that the Rashba spin-orbit field texture in such structures can be controlled by an electric field, allowing to tune it from radial to tangential. Such spin-orbit engineering should be useful for designing novel spin-charge conversion and spin-orbit torque schemes, as well as for controlling correlated phases and superconductivity in van der Waals materials.
\end{abstract}

\maketitle

\textit{Introduction.} The Rashba effect is the appearance of an extrinsic spin-orbit coupling (SOC) at surfaces and interfaces of electronic materials. Following the study by Rashba and Sheka on wurtzite semiconductors~\cite{Rashba1959:FTT}, the effect was formulated by Bychkov and Rashba for 2D electron gas~\cite{Bychkov1984:JETP}; see the recent review~\cite{Bihlmayer2022:NRP}. The Rashba effect is at the heart of spintronics, allowing for efficient spin manipulation in a variety of spin transport, spin relaxation, and spin detection phenomena~\cite{Zutic2004:RMP, Fabian2007:APS, Han2014,Avsar2020:RMP,Perkins2024}. 

The advent of two-dimensional (2D) materials and van der Waals (vdW) heterostructures has significantly expanded the range of possibilities for controlling the electron spin~\cite{Han2014,Avsar2020:RMP,Perkins2024}. In particular, novel spin interactions can be generated in vdW stacks by the proximity effect~\cite{Sierra2021:NN, Zutic2019:MT}. The prime example is the valley-Zeeman (Ising) SOC, induced in graphene (Gr) from a neighboring transition-metal dichalcogenide (TMDC) such as MoSe$_2$ or WSe$_2$, yielding a giant spin relaxation anisotropy~\cite{Cummings2017:PRL,Ghiasi2017:NL,Benitez2018:NP} and with controllable spin precession by gate voltages in spin transistor devices at room temperature~\cite{Ingla2021:PRL}.

In addition to the valley-Zeeman term, the breaking of space and mirror symmetries leads to the Rashba spin-orbit field (SOF). Pioneering tight-binding studies of twisted Gr/TMDC~\cite{Li2019, David2019} proposed that the Rashba coupling acquires a so-called Rashba angle $\varphi$ (between the electron's momentum and spin). The general form of the Rashba coupling at $K$ and $K'$ points for ${\cal{C}}_3$ symmetric systems is
\begin{equation}
\mathcal{H}_{\textrm{R}} = \lambda_{\textrm{R}}\textrm{e}^{-\textrm{i}\varphi s_z/2}(\tau \sigma_x \otimes s_y + \sigma_y \otimes s_x)\textrm{e}^{\textrm{i}\varphi s_z/2}.
\end{equation}
The coupling constant $\lambda_R$ denotes the strength of the Rashba field, $\tau = \pm 1$ is the valley index, $\sigma$ are the pseudospin and $s$ spin Pauli matrices. If $\varphi = 0$, the Rashba field is tangential (conventional); if $\varphi = 90$\degree, it is radial (unconventional). 

The Rashba angle $\varphi$ has been calculated for twisted Gr/TMDC from first-principles~\cite{Naimer2021:PRB, Zollner2023:PRB, Pezo2021:2DM, Lee2022:PRB} and by tight-binding modeling~\cite{Peterfalvi2022:PRB}, as well as for Gr/1T-TaS$_2$~\cite{Szalowski2023:2DM}. DFT calculations for twisted Gr/TMDCs~\cite{Naimer2021:PRB, Zollner2023:PRB, Lee2022:PRB} find that the Rashba angle varies between -20\degree\ and 40\degree, not being radial at any twist angle. Remarkably, the predicted variation of the Rashba angle has now been seen experimentally in Gr/WSe$_2$ bilayers~\cite{yang2023twistangle}, observing $\varphi$ up to about $\pm 60$\degree.

Here, we show that a radial Rashba SOF emerges in twisted homobilayers and multilayers of hexagonal lattices. Specifically, we perform first-principles calculations on four distinct structures: twisted bilayer graphene (TBLG), twisted bilayer WSe$_2$, twisted multilayers WSe$_2$/Gr/WSe$_2$, and WSe$_2$/TBLG/WSe$_2$. We show that in all the investigated cases the in-plane Rashba field can be radial, $\varphi = 90$\degree, but we also discuss cases in which it is not. For computational reasons, we use the twist angle of 21.79\degree\ (and the complementary 38.21\degree) for the homobilayers, noting that commensurate moiré crystals of bilayer WSe$_2$ twisted at these angles were recently created~\cite{Li2024Nature}. For such structures, our findings make even quantitative predictions. The origin of the radial Rashba field can be traced to the interference of two layer-locked (in untwisted stacks hidden) Rashba fields that have opposite tangential, but the same radial components. Important, the Rashba field can be tuned from radial to tangential by a displacement field.

Engineering radial Rashba coupling would allow to achieve unconventional charge-to-spin conversion in vdW heterostructures~\cite{Zhao2020:AM, Veneri2022:PRB, Lee2022:PRB, Ingla-Aynes2022:2DM, Safeer2019:NL2, Ontoso2023:PRA,Zollner2023:PRB}, improve the functionality of spin-orbit torque~\cite{Han2021:APL} by allowing to control the polarization of accumulated spin accumulation and spin current, and influence correlated phases~\cite{Lin2021, Xie2023:PRB, Zhumagulov2023_1:arxiv, Zhumagulov2023_2:arxiv, Koh2023:arxiv} and superconductivity~\cite{Zhang2023:N, Jimeno-Pozo2023:PRB, Holleis2023:arxiv}. 
Furthermore, a radial spin texture in a vdW heterostructure would emulate, in a controlled way, the parallel spin-momentum locking in chiral materials~\cite{Gosalbez-Martinez2023:PRB} such as 3D tellurium~\cite{Sakano2020:PRL, Calavalle2022:NM} or chiral topological semimetals~\cite{Bradlyn2016:S, Krieger2022:arxiv}. Very recently, chiral-induced spin selectivity 
has been predicted to be realized in vertical tunneling by twisted TMDCs \cite{Menichetti2024:P}. Our first-principles calculations show that, while not being universal, the formation of a radial Rashba spin texture can be readily achieved in twisted vdW structures.

\paragraph{Twisted bilayer graphene.} 
Monolayer Gr exhibits intrinsic spin-orbit splitting of tens of $\mu$eVs ~\cite{Gmitra2009, Sichau2019:PRL}, and no Rashba coupling; a transverse electric field induces a Rashba splitting of about 10 $\mu$eV per each V/nm ~\cite{Gmitra2009}. Similar values appear in bilayer graphene (BLG)~\cite{Konschuh2012, Banszerus2020}. 

The electronic structure of BLG with SOC
can be described as comprising conventional layer-dependent Rashba SOF with opposite orientations in the two layers~\cite{Konschuh2012}; the net result is no overall Rashba field and no spin-orbit splitting of the bands. In effect, each layer exerts an effective electric field on the other layer, causing a Rashba coupling in it. This is similar to the concept of ``hidden" spin polarization in inversion symmetric materials~\cite{Zhang2014:NP}. A transverse electric field induces layer polarization and removes the perfect balance of the hidden Rashba fields, giving a non-zero \emph{conventional} Rashba spin texture ~\cite{Konschuh2012}. Chirally stacked (ABC) graphene trilayers follow the same SOC trends~\cite{Kormanyos2013, Zollner2022:PRB}.

How do the hidden, layer-dependent Rashba fields manifest themselves in the electronic structure when the two layers get twisted? Without layer polarization, the two spin vector fields in the momentum space add up; if the fields of the hybridizing Bloch states from each layer are similar, these two tangential (but rotated) fields yield a purely radial, \emph{unconventional} Rashba spin texture. Layer polarization---due to a transverse electric field or the presence of a substrate---can add a tangential component and make the Rashba field fully in-plane tunable. 

\begin{figure}
  \includegraphics[width=0.95\columnwidth]{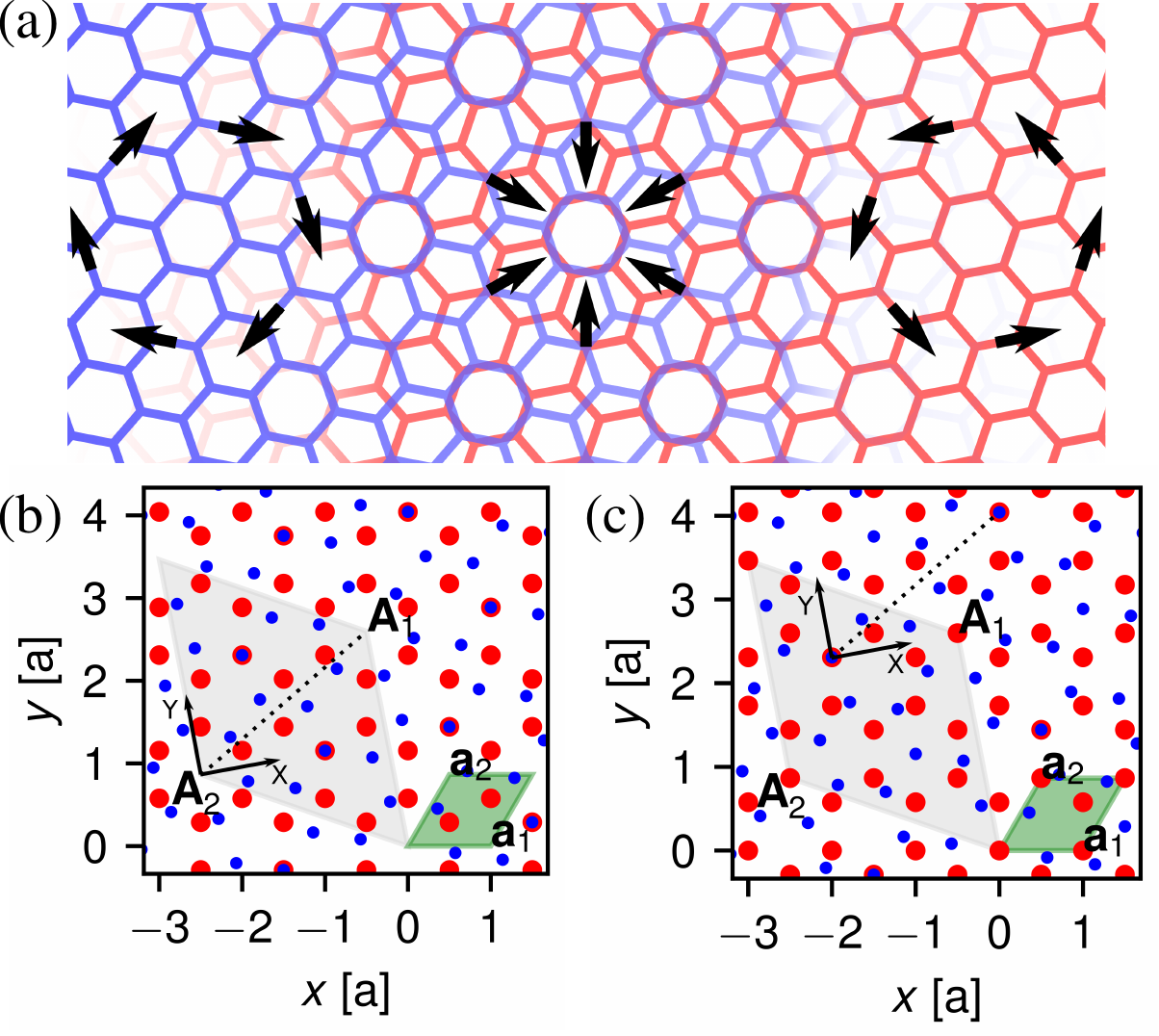}
  \caption{(a) Sketch of radial Rashba emergence by the interference of oppositely rotating tangential Rashba fields. (b,c) 21.79\degree\ TBLG ($p$=1, $q$=3) and different sublattice-exchange symmetries. The primitive cell of graphene with its lattice vectors $\vec{a}_{1/2}$ is indicated by the green rhombus. The shaded area is the moir\'{e} supercell with lattice vectors $\vec{A}_{1/2}$. The red/blue dots represent carbon atoms from the bottom/top layer. The even system (b) is obtained from AA stacking and rotation about the origin $[0,0]$, the odd system (c) from initial AB (Bernal) stacking. The coordinate system for DFT calculations is shown by labels X and Y. In-plane twofold rotation axes are indicated by dashed lines.}
  \label{fig:cells}
\end{figure}

We illustrate this concept by performing first-principles simulations and effective modeling for 21.79\degree\ (sublattice-exchange) even and odd TBLG ~\cite{Mele2010}, which exhibit three-fold rotational symmetry and resemble energy-renormalized versions of AA and AB (Bernal) BLG dispersions, respectively. The supercells are constructed following Shallcross \textit{et al}.~\cite{Shallcross2013}. Further geometry setup details are given in \cite{supp}. Even TBLG has two overlapping atomic sites, see Fig.~\ref{fig:cells}(b), whereas the odd structure has only one eclipsed position, see Fig.~\ref{fig:cells}(c). 
In twisted structures, spatial modulation of the interlayer interaction provides momentum conservation for the coupling of different single-layer momentum states via Umklapp processes~\cite{Mele2010}. The interaction between the layers predominantly happens at the eclipsed sites and the reduction in the energy scale (with respect to BLG analogues) is a measure for the loss of interlayer registry~\cite{Mele2011}.

\begin{figure}	 
  \includegraphics{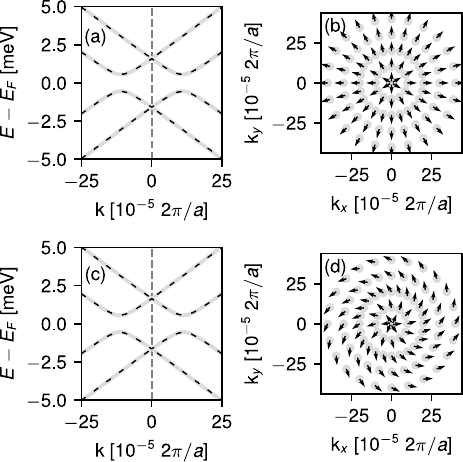}
  \caption{\emph{Ab initio} low-energy band structures and spin-fields of 21.79\degree\ \textit{even} TBLG. (a) Low-energy TBLG band structure, (b) SOF of the highest valence band, (c) TBLG band structure with an electric field 29 mV/nm, and (d) SOF of the highest valence band in the electric field. The vector field encodes $s_x$ and $s_y$ components of the spin expectation value. The K point is marked by the vertical dashed line. The color code denotes the spin-z expectation value, where red is spin-up, grey color denotes zero polarization and blue is spin-down. The dashed lines on top of the DFT bands are model fits with parameters $v_\text{F} = 8.16 \cdot 10^5$m/s, $w=1.597$~meV, $\LI=-11.858~\mu$eV, $\LR=13.5~\mu$eV, and $u=0.43$~meV.}
  \label{fig:band_spin_even}
\end{figure}

The band structures of even and odd TBLGs are calculated by the \texttt{Wien2k} code\cite{Blaha2020}, which accounts for $d$ orbitals responsible for SOC in Gr~\cite{Gmitra2009} (for computational details see the Supplemental Material~\cite{supp}). The band structure and spin texture of the even system are shown in Fig.~\ref{fig:band_spin_even}. We focus on the low-energy physics at the K point of the moir\'{e} lattice, indicated by the meV energy scale. The band structure exhibits two crossings at the K point, which are remnants of the Dirac cones of single-layer graphene separated in energy by 3.6~meV. Shifting the cones would resemble the rescaled AA-stacked BLG band structure~\cite{Mele2010,supp}. However, the interlayer interaction opens a gap of 1.1~meV between the two copies of Dirac cones. The Dirac cones themselves are gapped (not visible) by a spin-orbit gap of 24~$\mu$eV as in single layer Gr~\cite{Gmitra2009}. The spin degeneracy is lifted by $\sim10~\mu$eV due to missing inversion symmetry. The spin expectation values of the highest valence band around K, shown in Fig.~\ref{fig:band_spin_even}(b), form a purely radial Rashba SOF. All the low-energy bands have such a spin texture, with alternating directions pointing towards or away from the K point. However, the radial texture can be tuned to tangential by applying an out-of-plane electric field, as shown in Fig.~\ref{fig:band_spin_even}(d) for the field of 29~mV/nm, while the band structure is hardly affected, see Fig.~\ref{fig:band_spin_even}(c).  

\begin{figure}
  \includegraphics{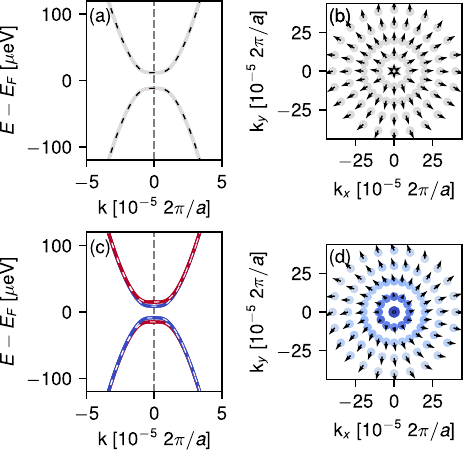}
  \caption{Same as Fig.~\ref{fig:band_spin_even} but for 21.79\degree-rotated \textit{odd} TBLG in an electric field of 29~$\mu V$/nm. Model fit parameters are $v_\text{F} = 8.16 \cdot 10^5$m/s, $w=1.685$~meV, $\phi=0.171$, $\LI=-11.858~\mu$eV, $\LR=0.3~\mu$eV, and $u=3.1~\mu$eV.}
  \label{fig:band_spin_odd}
\end{figure}

In the case of the odd system, Fig.~\ref{fig:band_spin_odd}, the band structure has a quadratic dispersion, similar to AB-stacked BLG. The valence-conduction band degeneracy at K is lifted by the intrinsic spin-orbit gap of 23~$\mu$eV and band splittings on the sub-$\mu$eV scale are introduced. The parabolic second valence and conduction bands, see e.g. Ref.~\onlinecite{Mele2010}, are split off by 1.7~meV, outside the energy window. In the odd system we recognize a radial shape of the spin texture as well. The SOF shows some deviations from pure radial, because the symmetry is reduced from sixfold rotation to a threefold symmetry. Application of a tiny electric field of 29~$\mu$V/nm further opens the gap and the band splitting saturates at the intrinsic SOC energy scale. We find that the band structure becomes spin-polarized and that the spin texture acquires a tangential component.

The calculated low-energy bands along with the spin textures can be quantitatively described by an effective model of TBLG with layer-dependent intrinsic and (hidden) Rashba couplings. We refer to~\cite{supp} for details of the model, but present in Figs.~\ref{fig:band_spin_even} and \ref{fig:band_spin_odd} the fits.  

We note that our calculated spin textures are at odds with the recent report~\cite{Yananose2021} which considered the same twist angle of TBLG, but found vortex-like spin textures at $K$, employing  \texttt{VASP} code~\cite{Kresse1996:PRB}.
To crosscheck our results, we also employed \texttt{Quantum Espresso} ~\cite{Giannozzi2009:JPCM} and confirmed the radial spin textures for both unrelaxed and relaxed geometries, without significant differences. Also, our effective model \cite{supp} fits well the DFT simulations, see Figs.~\ref{fig:band_spin_even} and \ref{fig:band_spin_odd}, giving additional support for the emergence of radial Rashba fields. 

\begin{figure}
  \includegraphics{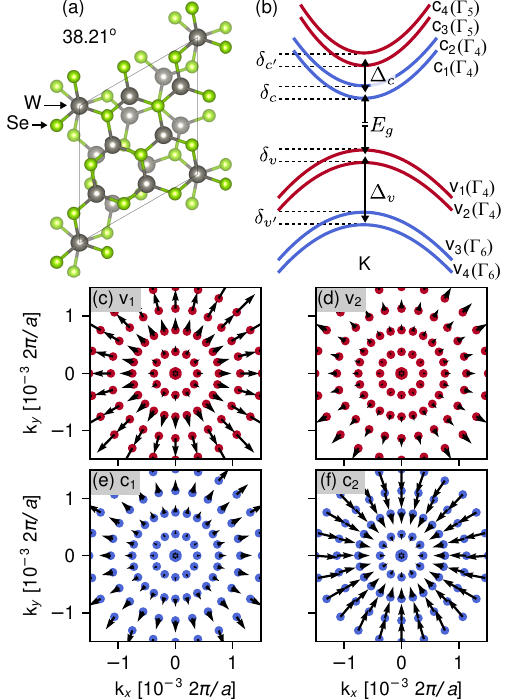}  
  \caption{Radial Rashba in twisted WSe$_\text{2}$ homobilayers. (a) 38.21\degree\ supercell with W and Se atoms indicated. (b) Sketch of the low-energy bands indicating the relevant energy splittings and their irreducible representations. The calculated \textit{ab initio} splittings (in meV) are $E_g = 1352.79$, $\Delta_v = 453.22$, $\Delta_c = 40.65$, $\delta_{v^\prime} = 4.29$, $\delta_{v} = 1.94$, $\delta_{c} = 0.63$, $\delta_{c^\prime} = 0.24$. Calculated spin textures for the energy bands (c) v$_1$, (d) v$_2$, (e) c$_1$ and (f) c$_2$. The radial Rashba texture points outwards for v$_1$, v$_2$, and c$_1$ and inwards for c$_2$.}
  \label{fig:twistedWSe2}
\end{figure}

\textit{Twisted WSe$_\text{2}$ homobilayers.} To demonstrate that not only TBLG exhibits purely radial Rashba SOFs, we performed first-principles simulations of a twisted WSe$_2$ homobilayer using the \texttt{Wien2k} code~\cite{Blaha2020}. Monolayer TMDCs such as WSe$_\text{2}$ lack space inversion symmetry, so their electronic states are naturally spin split. However, the presence of a horizontal mirror plane symmetry precludes the appearance of in-plane Rashba fields but rather enables robust spin-polarization in the out-of-plane direction\cite{Xiao2012PRL}. Conversely, naturally stacked bilayer TMDCs have space inversion symmetry and no spin-orbit polarization of their bands. The twisted structures start from a 0\degree\ stacking with W and Se atoms on top of each other (the R$^{\text{h}}_{\text{h}}$ stacking\cite{Tran2019Nature, FariaJunior2023Nanomat}, containing a horizontal mirror plane). We discuss here in the main text the commensurate unit cell for a twist angle of 38.21\degree\ with the corresponding $C_3$ symmetric atomic structure shown in Fig. \ref{fig:twistedWSe2}(a). 

\begin{figure}
  \includegraphics[width=0.99\columnwidth]{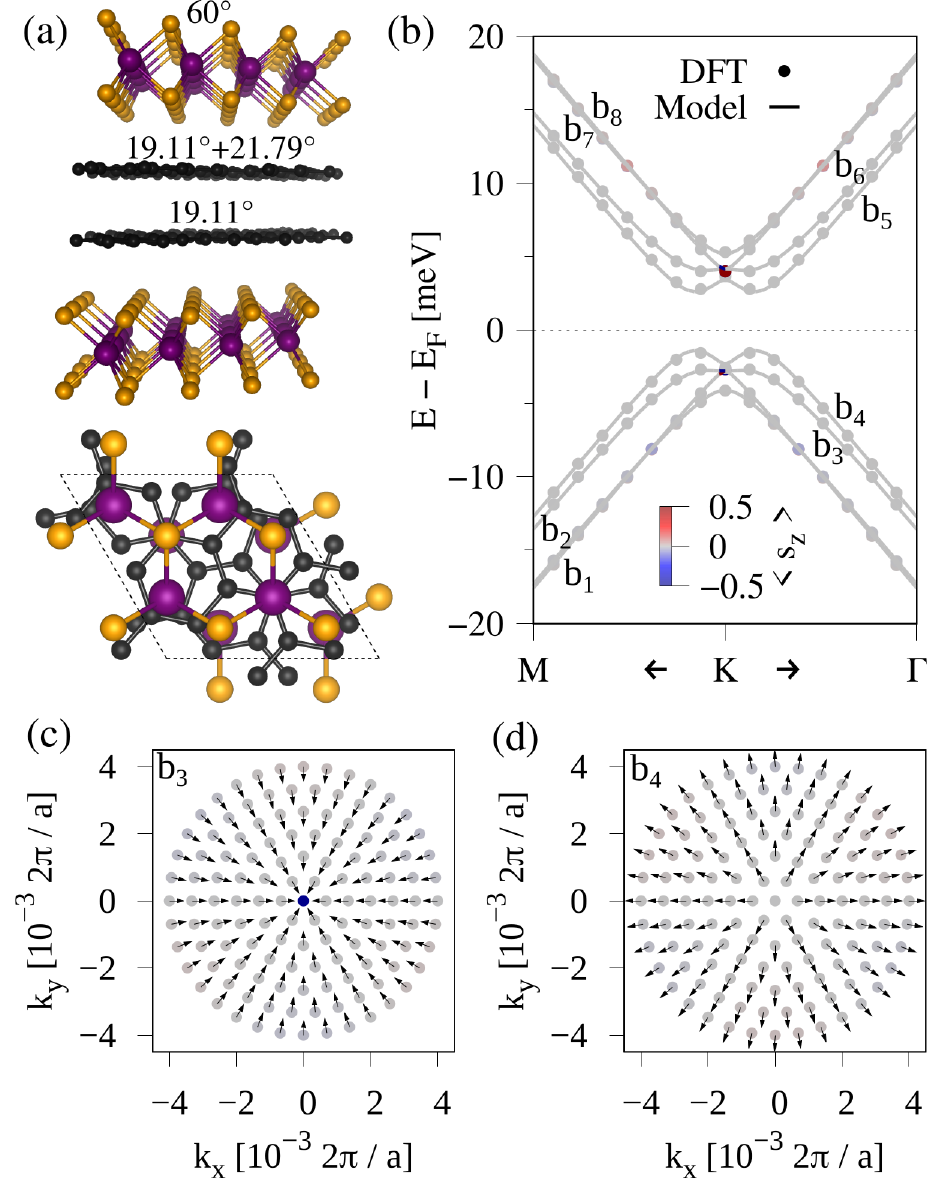}
  \caption{\label{fig:TBLG_WSe2_encap_main} Radial Rashba in WSe$_2$-encapsulated even TBLG. (a) Top and side view of the twisted multilayer stack. The twist angle indicated above each layer is measured with respect to the bottom WSe$_2$ layer. (b) The corresponding low-energy bands with a fit to the model Hamiltonian. (c) and (d) Exemplary spin textures of the bands b$_3$ and b$_4$ as labeled in subfigure (b).}
\end{figure}

The band structure of the 38.21\degree\ supercell is presented in Fig.~\ref{fig:twistedWSe2}(b), indicating monolayer- and moiré-derived
spin-orbit splittings, $\Delta$ (consistent with isolated monolayers\cite{Kormanyos2015TDM}) and 
$\delta$, respectively. The spin textures for the lowest energy bands, v$_1$-v$_2$ and c$_1$-c$_2$, are given in Figs.~\ref{fig:twistedWSe2}(c-f). Despite the strong spin-valley locking (out-of-plane spins)\cite{Xiao2012PRL}, our calculations clearly reveal the emergence of in-plane radial Rashba textures in the vicinity of the K-valleys. The corresponding in-plane spin expectation values are on the order of $10^{-5} - 10^{-4}$, roughly 3 orders of magnitude smaller than in Gr systems (see Fig.~S11) but well above the numerical precision\cite{Kurpas2019PRB}. In the Supplemental Material~\cite{supp}, we show the full band structure and the spin textures for all the valence and conduction bands v$_1$-v$_4$ and c$_1$-c$_4$, as well as the same analysis for the complementary twist angle of 21.79\degree. Particularly, in the 21.79\degree\ case the radial Rashba spin textures acquire trigonal features that become more pronounced as we move away from the K-valleys for particular energy bands, due to the hybridization of the Bloch states from different layers occurring at different points in the Brillouin zone. These distinct k-points can exhibit hidden Rashba fields with unequal magnitudes and/or angular textures (see Fig.~S7 for monolayer WSe$_\text{2}$ under electric field).

\textit{WSe$_\text{2}$-encapsulated TBLG.}
BLG has weak Rashba coupling, but encapsulated by TMDCs the spin-splitting of BLG is on the meV scale, well within the reach of spin transport experiments. Here, we demonstrate the existence of unconventional Rashba coupling in twisted multilayer stacks, containing either monolayer Gr (see~\cite{supp}) or TBLG, see Fig.~\ref{fig:TBLG_WSe2_encap_main}. The twist angle of BLG is 21.79\degree, while the bottom layer of BLG is twisted by 20.11\degree\ from the adjacent WSe$_2$. To complete the chiral structure, the top WSe$_2$  monolayer is twisted by 60\degree\ with respect to the bottom WSe$_2$; see Fig.~\ref{fig:TBLG_WSe2_encap_main}(a). 

The energetically spin-split electronic states have no out-of-plane spin orientation, as shown in Fig.~\ref{fig:TBLG_WSe2_encap_main}(b). The in-plane spin polarizations, presented in Fig.~\ref{fig:TBLG_WSe2_encap_main}(c) and (d), are purely radial. The difference to BLG is the magnitude of the Rashba coupling, which is about 1 meV. We fit the DFT data to a model Hamiltonian in Supplemental Material\cite{supp}. The Hamiltonian description should be useful for investigating spin transport and correlation physics in such chiral multilayers. Important, the spin texture can turn from radial to tangential, upon applying an out-of-plane electric field~\cite{supp}.

\textit{Conclusions.} 
We studied the emergence of purely radial SOFs in twisted homobilayers of Gr and WSe$_2$, as well as in twisted multilayer heterostructures comprising Gr and WSe$_2$. We found that this unconventional Rashba spin texture is fully in-plane tunable, from radial to tangential, by a displacement field. Such SOFs should be even more pronounced in homobilayers of strong spin-orbit materials with built-in  Rashba SOC, such as Janus dichalcogenides~\cite{Soriano2021:NJP} as well as in vdW heterostructures with strong interlayer coupling which induces strong hidden Rashba fields. 

\begin{acknowledgments}
The authors thank Martin Gmitra, Denis Kochan, and Thomas Naimer for fruitful discussions. This work was funded by the Deutsche Forschungsgemeinschaft (DFG, German Research Foundation) SFB 1277 (Project-ID 314695032). The authors gratefully acknowledge the Gauss Centre for Supercomputing e.V. (www.gauss-centre.eu) for funding this project by providing computing time on the GCS Supercomputer SuperMUC at Leibniz Supercomputing Centre (LRZ, www.lrz.de).
\end{acknowledgments}

\bibliography{paper}

\end{document}


\title{Emergence of radial Rashba spin-orbit fields in twisted van der Waals heterostructures: Supplemental Material}

\author{Tobias \surname{Frank}}

\author{Paulo E. \surname{Faria~Junior}}
\email[Email to: ]{fariajunior.pe@gmail.com}

\author{Klaus \surname{Zollner}}

\author{Jaroslav \surname{Fabian}}

\affiliation{Institute for Theoretical Physics, University of Regensburg, 93040 Regensburg, Germany}
 
\date{\today}

\maketitle

\tableofcontents

\clearpage

\section{Twisted bilayer graphene (TBLG)}

\subsection{Computational details and geometry setup}

The supercells are constructed following Shallcross \textit{et al}.~\cite{Shallcross2013}.
 We define the primitive graphene cell as $\vec{a}_1 = (1, 0)_\mathrm{xy}$ and $\vec{a}_2 = (-1/2, \sqrt{3}/2)_\mathrm{xy}$ and atoms at positions $(1/3, 1/3)_{12}$ and $(2/3, 2/3)_{12}$.
The supercell lattice vectors take the form {$\vec{A}_1 = [-(p + q) \vec{a}_1 + 2 q \vec{a}_2] / \gamma$ and $\vec{A}_2 = [-2q \vec{a}_1 - (p - q) \vec{a}_2] / \gamma$},
with $\gamma = \gcd(p+3q, p-3q)$. For our cells, we choose $(p, q) = (1, 3)$, resulting in an rotation angle of 
\begin{equation}
	\theta = \tan^{-1}[(3(p/q)^2 - 1)/(3(q/p)^2 + 1)] \approx 21.79^{\circ}. \label{twist_angle}
\end{equation}
The layer lattice vectors in the bottom (b) and top (t) layers in terms of the bare Gr lattice vectors are given by {$\vec{R}_b = i \vec{a}_1 + j \vec{a}_2 + \vec{t}$ and $\vec{R}_t = R_\theta (i \vec{a}_1 + j \vec{a}_2)$ with integer $i$ and $j$.}
Translation vector $\vec{t}$ controls the sublattice-exchange symmetry: the even system results from $\vec{t}_\mathrm{even} = (0, 0)_{12}$,  the odd from  $\vec{t}_\mathrm{odd} = (1/3, 1/3)_{12}$. The top layer is rotated with respect to the bottom layer by the rotation matrix $R_\theta$, taking the angle of Eq.~\eqref{twist_angle} to ensure commensurability. Our even and odd structures are shown in Fig.~1. 
 The DFT code we employ, \texttt{Wien2k}
 ~\cite{Blaha2020}, symmetrizes the supercell internally such that $\vec{A}_1$ is aligned with the $y$ direction; this is important to correctly interpret 
 in-plane spin components.

The resulting twisted bilayers lack inversion symmetry. The even system belongs to space group P622, having the $z$-axis as a six-fold rotation axis and a two-fold rotation symmetry around the axis indicated in Fig.~1(b), giving a total of 12 symmetry operations. The odd system has half the number of symmetry operations and its space group is P321 with a threefold rotation symmetry about the $z$-axis and a two-fold in-plane rotation axis.

Our electronic structure calculations within the LAPW (linearized augmented plane wave) formalism are carried out with \texttt{Wien2k}~\cite{Blaha2020}, an electronic structure code that gives accurate results for SOC in carbon-based materials~\cite{Gmitra2009, Konschuh2012}. The lattice constant of graphene is set to $a=2.46$~\AA~and a vacuum spacer of 12~\AA~is inserted between repeated structures in $z$-direction of our slab calculation. We apply a $10$ \AA$^{-1}$ cutoff for plane waves and employ a $k$-point grid of $6\times 6$. Empirical van der Waals corrections~\cite{Grimme2010:JCP} lead to an interlayer distance of $3.49$~\AA, in agreement with accurate quantum Monte Carlo calculations for BLG~\cite{Mostaani2015}. 

In the paper, we show the results for unrelaxed TBLG. We have checked, using \texttt{Quantum ESPRESSO}, that relaxing the structures does not alter the radial spin textures. The relaxation of the structures changes only slightly the 120\degree~angles by 0.05\degree, and introduces a rippling of about 0.003 $\textrm{\AA}$.

\subsection{AA stacked bilayer graphene}

\begin{figure}[]
    \includegraphics[width=0.8\textwidth]{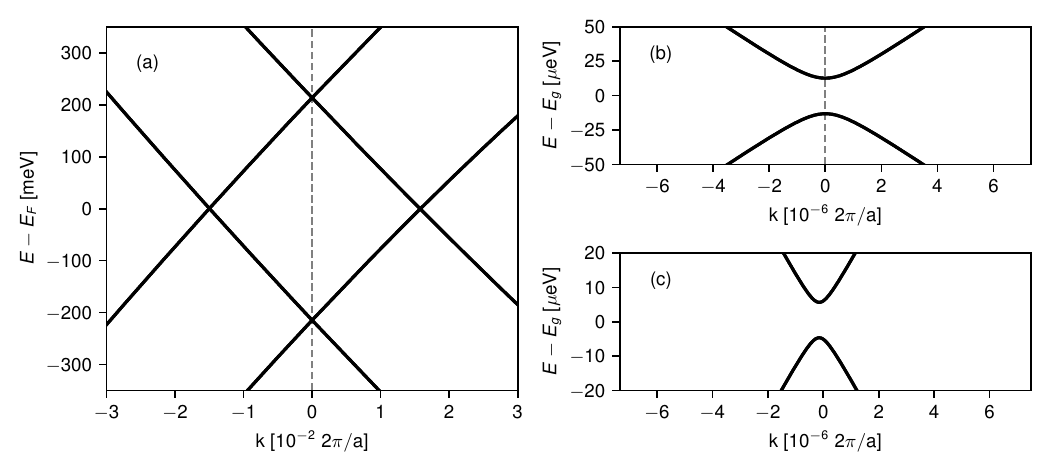}
    \caption{DFT-calculated AA-stacked bilayer graphene. (a) global band structure, (b) zoom into the crossing at K (k=0, E=200 meV) -- intrinsic SO gap, (c) zoom into the crossing off K -- SOC-mediated gap (k =1.5, E = 0)}
    \label{fig:shift}
\end{figure}

The even TBLG derives from the energetically less favourable AA stacking. For completeness, we also present the \textit{ab initio} calculation of the low-energy energy bands of AA BLG in the presence of SOC, see Fig.\ref{fig:shift}.

The band structure consists of two Dirac cones split by the interlayer coupling. Both the valence and conduction Dirac bands feature spin-orbit (intrinsic) gaps of about 25 $\mu$eV. The valence and conduction bands exhibit SOC-induced anticrossings off $K$.

\subsection{Effective Hamiltonian for TBLG with spin-orbit coupling.}
\label{sec:models}

The emergence of a purely radial Rashba spin-orbit field in graphene is unprecedented and the mechanism leading to it deserves special examination. To this end, we extend the low-energy Hamiltonians for large-angle TBLG of Refs.~\onlinecite{Mele2010} and \onlinecite{Weckbecker2016} with the corresponding SOC terms. TBLG can be modeled by expressing the Hamiltonian in terms of single-layer graphene Bloch functions~\cite{Bistritzer2011, Weckbecker2016, Mele2010}, applying a cutoff at the momentum coupling scale $g=8\pi/\sqrt{3}a\sin{\theta/2}$. Weckbecker et al.~\cite{Weckbecker2016} provide a unified description of large and small angle regimes and find agreement with tight-binding calculations. Our \textit{ab initio} results correspond to the electronic structure given in Refs.~\onlinecite{Mele2010} and~\onlinecite{Weckbecker2016}.

The orbital Hamiltonians read\cite{Weckbecker2016, Mele2010}
\begin{equation}
	H_\text{even}(\vec{k}) = \begin{pmatrix}
	u & \hbar v_F k e^{i (\theta_k - \theta/2)} & w e^{-i\phi} & 0\\
	\hbar v_F k e^{-i (\theta_k - \theta/2)} & u & 0 & we^{i\phi}\\
        w e^{-i\phi} & 0 & -u & \hbar v_F k e^{i (\theta_k + \theta/2)}\\
        0 & we^{i\phi} & \hbar v_F k e^{-i(\theta_k + \theta/2)} & -u
	\end{pmatrix} \label{eq:even_hamiltonian}
\end{equation}
and
\begin{equation}
	H_\text{odd}(\vec{k}) = \begin{pmatrix}
        u & \hbar v_F k e^{i (\theta_k - \theta/2)} & w & 0\\
        \hbar v_F ke^{-i (\theta_k - \theta/2)} & u & 0 & 0\\
        w & 0 & -u & -\hbar v_F k e^{-i (\theta_k + \theta/2)}\\
        0 & 0 & -\hbar v_F k e^{i (\theta_k + \theta/2)} & -u
	\end{pmatrix} \label{eq:odd_hamiltonian}.
\end{equation}

The outer blocks describe the layer degree of freedom (the first block is the bottom layer), and the inner basis is the sublattice degree of freedom. The linearized graphene dispersions with Fermi velocity $v_F$ within the layers are symmetrically rotated by the relative angle $\theta$. The angles $\theta$ and $\theta_k=\arg(k_x+ik_y)$ are measured with respect to the $k_x$ axis. The interlayer coupling $w$ reflects effective coupling within a single sublattice in the odd system. Different signs in the arguments of the exponentials and prefactors indicate that in the odd case, K and K$^\prime$ from different layers are coupled\cite{Mele2010}. Our notation follows Ref.~\onlinecite{Kochan2017}, to have consistent graphene [$\hbar v_F (\kappa k_x\sigma_x-k_y\sigma_y)$] and SOC Hamiltonians. The valley is addressed by $\kappa$, which takes values of $\pm 1$ for K and K$^\prime$. Angle $\phi$ is added as a free parameter in Eq.~\eqref{eq:even_hamiltonian}~\cite{Mele2010}, necessary to tune the anticrossing in the even bilayer system. To model the electric field of our \textit{ab initio} calculations, we put potentials of $\pm u$ onto the bottom and top layers to simulate an out-of-plane electric field (positive u -- positive field in $z$), respectively. 

Knowledge about which K points are coupled is important for the SOC extension. In graphene, the low-energy Rashba SOC Hamiltonian is given by $\lambda_R(-\kappa\sigma_x s_y - \sigma_y s_x)$, where $s$ and $\sigma$ Pauli matrices indicate the real spin and sublattice degree of freedoms. Hence, the Rashba SOC extension to Hamiltonians of Eqs.~\eqref{eq:even_hamiltonian} and \eqref{eq:odd_hamiltonian} are
\begin{align}
	 H_{\text{even, }R} =
	\begin{pmatrix}
		\lambda_R(-\sigma_x s_y - \sigma_y s_x) & 0 \\
        	0 & -\lambda_R(-\sigma_x s_y - \sigma_y s_x)
    	\end{pmatrix},
\end{align}
and
\begin{align}
	 H_{\text{odd, }R} =
	\begin{pmatrix}
		\lambda_R(-\sigma_x s_y - \sigma_y s_x) & 0 \\
        	0 & -\lambda_R(\sigma_x s_y - \sigma_y s_x)
    	\end{pmatrix}.
\end{align}

The signs of Rashba SOC in the bottom and top layer are opposite due to the crystal field gradients that point into opposite perpendicular directions, responsible for the Rashba effect.

The other ingredient for a minimal SOC model in bilayer graphene~\cite{Konschuh2012} is the intrinsic SOC, which is given by $\lambda_I(\kappa \sigma_z s_z)$, leading to the extension

\begin{align}
	H_{\text{even, }I} = 
	\begin{pmatrix}
		\lambda_I \sigma_z s_z & 0 \\
		0 & \lambda_I\sigma_z s_z
    	\end{pmatrix},
\end{align}
and
\begin{align}
	H_{\text{odd, }I} = 
	\begin{pmatrix}
		\lambda_I \sigma_z s_z & 0 \\
		0 & \lambda_I(-\sigma_z s_z)
    	\end{pmatrix}.
\end{align}

Our \textit{ab initio} band structures on the orbital scale are reproduced very well by the models, see Figs. 2 and 3 in the main text. The intrinsic SOC in graphene of $\lambda_I = 12~\mu$eV remains unchanged, Rashba SOC is introduced. In principle, an applied electric field also changes the size of Rashba SOC by 10~$\mu$eV per V/nm aligning the Rashba couplings~\cite{Gmitra2009}, which is too small for the electric fields considered here.

\begin{figure}
   \includegraphics{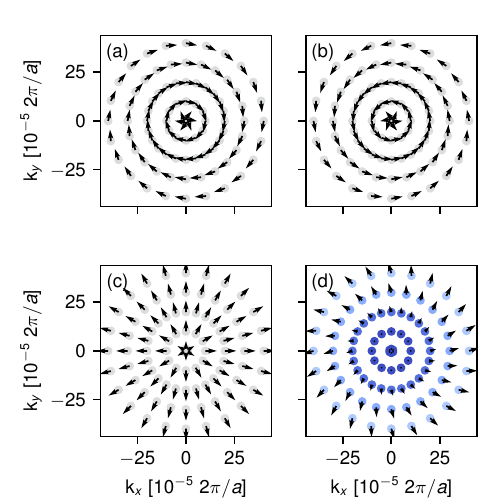}
   \caption{Model radial Rashba spin texture formation for the odd system. Panel (a) shows the spin field resulting from the isolated bottom-layer block of the model Hamiltonian, (b) from the top layer. In Panel (c), the spin field of the highest valence band is shown for the interacting system. In Panel (d), spin expectation values with respect to an electric field are displayed. The blue color indicates the strength of spin-down polarization gradually evolving to a grey color standing for zero spin-$z$.}
   \label{fig:model}
\end{figure}

The SOC-extended models for the even and odd systems can explain the emergence of radial Rashba. In Fig.~\ref{fig:model} we plot spin expectation values for the uncoupled layer blocks of the odd Hamiltonian. The bottom layer shows a spin field winding clockwise around the K point similar to the plain Rashba effect in graphene. Due to the rotation of the coordinate system with angle $\theta/2$, the spins in the $k$ points are not purely tangential to the equi-momentum circles but acquire a radial component. The spin field in the other layer has the opposite rotation sense due to the different signs in Rashba. The interaction between the layers leads to a vector addition of the spin expectation values in each $k$ point, which results in a cancellation of tangential components. This also means, that the spin texture is independent of the rotation angle and this effect should be observable in a range of large twist angles.

\subsection{Model Hamiltonian in electric field}

\begin{figure*}
  \includegraphics[width=\textwidth]{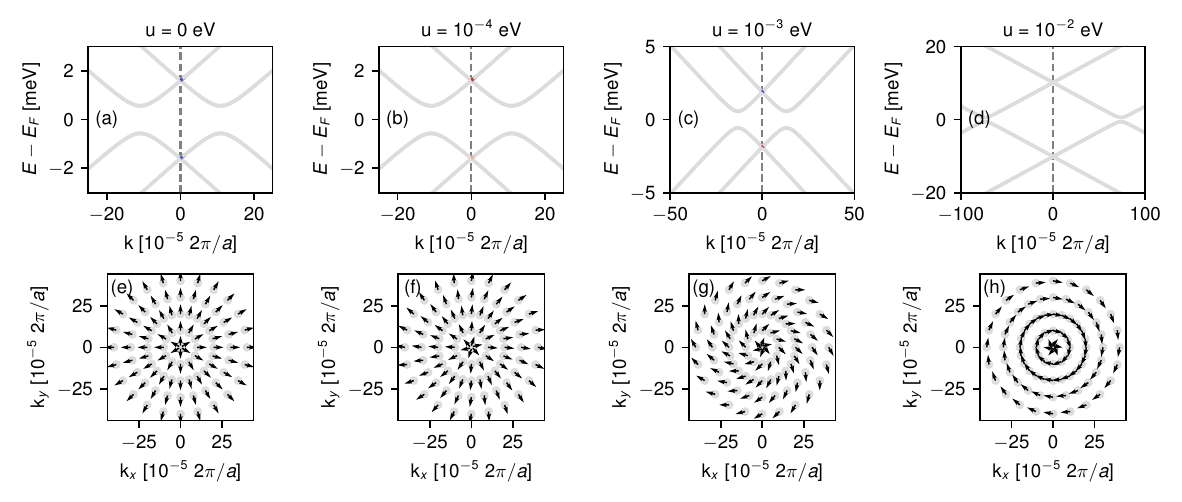}
  \caption{Electric field dependence of the even model system applying different potential differences $u$ across the layers. Model parameters as in main text.}
  \label{fig:21.79_AA_bands_spins_efield}
\end{figure*}

In Fig. \ref{fig:21.79_AA_bands_spins_efield}
we show the low-energy band structure of the even TBLG,  from the model Hamiltonian in the main text, as a function of the out-of-plane electric field. The off-$K$ anticrossings move to larger $k$ values as the field increases, but the overall topology of the bands remains the same. On the other hand, the spin texture is tuned from radial to tangential, demonstrating the full in-plane tunability of the Rashba field by the electric field.  

\begin{figure*}
  \includegraphics[width=\textwidth]{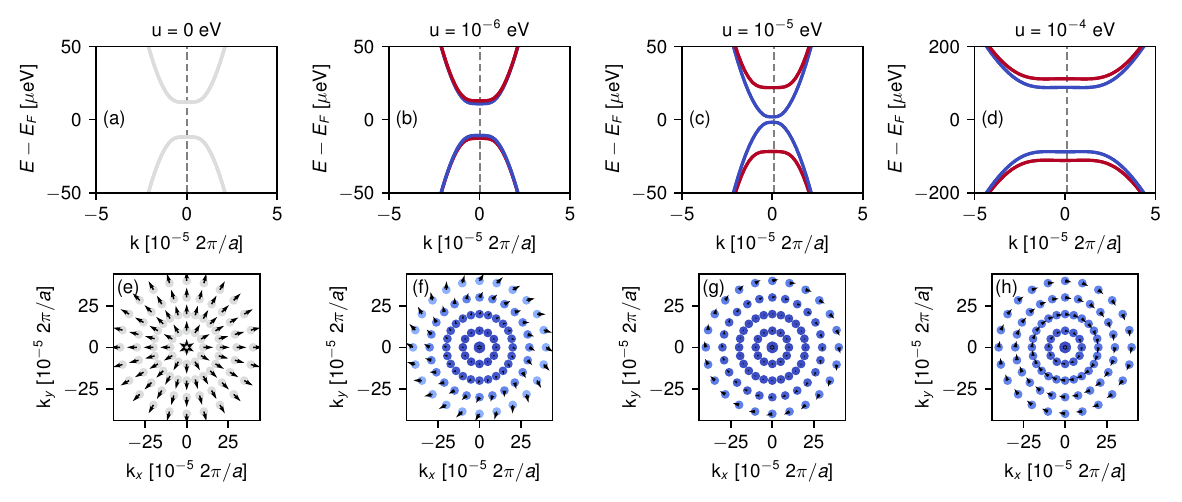}
  \caption{Electric field dependence of the odd model system applying different potential differences $u$ across the layers. Model parameters as in main text.}
  \label{fig:21.79_AB_bands_spins_efield}
\end{figure*}

In Fig. \ref{fig:21.79_AB_bands_spins_efield}
we show the electric field dependence of the electronic states in odd TBLG. The electric field effects are more interesting than for the even system. First, the electric field induces a sizeable (up to 25 $\mu$eV) spin-orbit splitting of the bands and closing the band gap, and rotates the spin texture from radial to tangential. As the electric field further increases, the spin-orbit splitting remains but the band gap reverses its trend and increases due to the layer polarization. The spin texture remains tangential.  

\cleardoublepage 

\section{Twisted WS$\text{e}_\text{2}$ homobilayers}

The \textit{ab initio} calculations of twisted WSe$_\text{2}$ homobilayers for two complementary twist angles, $\sim$21.79\degree~and $\sim$38.19\degree, are performed using \texttt{Wien2k}\cite{Blaha2020}. We employ the Perdew-Burke-Ernzerhof\cite{Perdew1996:PRL} exchange-correlation functional with van der Waals interactions included via the D3 correction\cite{Grimme2010:JCP}. We used a k-grid of $12 \times 12 \times 1$, and convergence criteria of 10$^{-5}$ $e$ for the charge and 10$^{-5}$ Ry for the energy. The plane-wave cutoff multiplied by the smallest atomic radii is set to 8. Spin–orbit coupling was included fully relativistically for core electrons, while valence electrons were treated within a second-variational procedure\cite{Singh2006} with the scalar-relativistic wave functions calculated in an energy window up to 5 Ry. The lattice parameter for the monolayer WSe$_\text{2}$ is $3.282 \; \textrm{\AA}$ and thickness is $3.34 \; \textrm{\AA}$\cite{FariaJunior2022NJP}. The interlayer distance is $3.4 \; \textrm{\AA}$\cite{Lin2021NatComm}. The vacuum is $20 \; \textrm{\AA}$. To create the twisted structures, the two WSe$_2$ are aligned at zero degree and the rotation axes goes through the W atoms (notice how the W atoms are on top of each other at the corners of the supercell in both cases), following Ref.~\cite{Uchida2014PRB}.

\begin{figure}[H]
    \includegraphics[width=\textwidth]{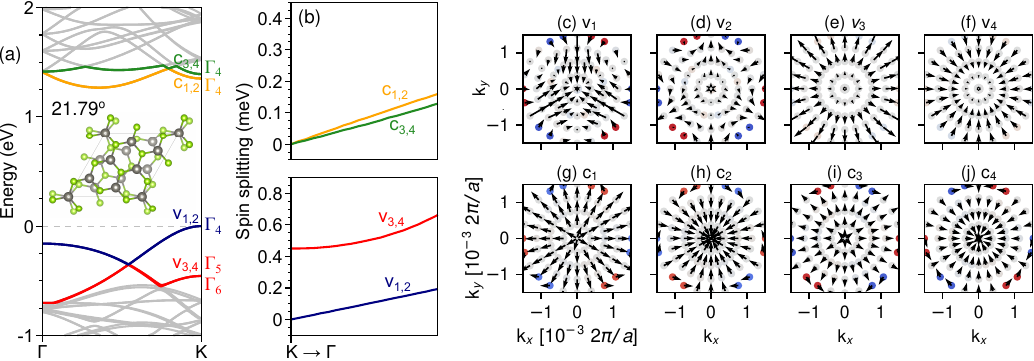}
    \caption{WSe$_\text{2}$ bilayer twisted by $\sim$21.79\degree. (a) Band structure along the $\Gamma$-K direction highlighting the low energy valence (v$_\text{1-4}$) and conduction (c$_\text{1-4}$) bands. The double group irreps at the K-point ($D_3$ point group) are indicated. The irrep $\Gamma_4$ is two-dimensional and real, while the irreps $\Gamma_5$ and $\Gamma_6$ are one-dimensional and complex. The inset shows the supercell. (b) Spin splitting of relevant low-energy bands around the K-point. (c)-(f) Spin texture around the K-point for the valence bands v$_\text{1}$-v$_\text{4}$. (g)-(j) Spin texture around the K-point for the conduction bands c$_\text{1}$-c$_\text{4}$.}
\label{fig:twistedWSe2_21.8}
\end{figure}

\begin{figure}[H]
    \includegraphics[width=\textwidth]{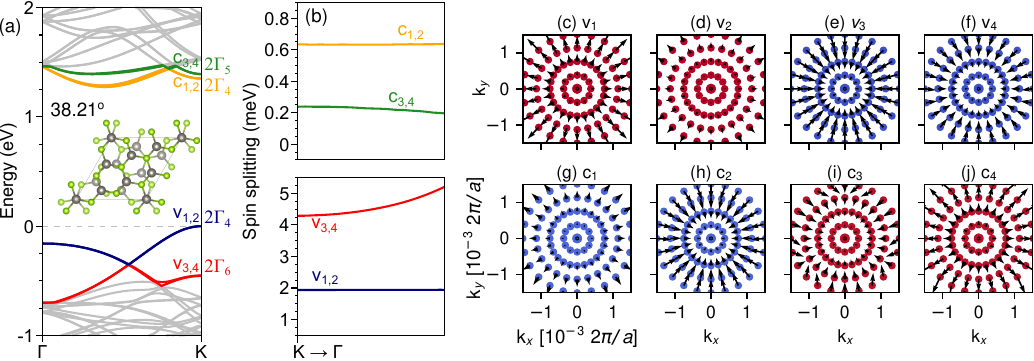}
    \caption{Same as Fig.~\ref{fig:twistedWSe2_21.8} but for the WSe$_\text{2}$ bilayer twisted by $\sim$38.21\degree. The point group at the K-point is $C_3$ with the double group irreps $\Gamma_4$ and $\Gamma_5$ being one-dimensional and complex, while the irrep $\Gamma_6$ is one-dimensional and real.}
\label{fig:twistedWSe2_38.2}
\end{figure}

The resulting band structures, spin splittings, and spin textures are presented in Figs.~\ref{fig:twistedWSe2_21.8} and \ref{fig:twistedWSe2_38.2} for $\sim$21.79\degree~and $\sim$38.19\degree~bilayer WSe$_\text{2}$ structures, respectively. We focus on the low energy valence (v$_\text{1-4}$) and conduction (c$_\text{1-4}$) bands. The $\sim$21.79\degree~structure belongs to the symmetry group $D^2_3$ (space group 150, P321), with its K-point belonging to the point group $D_3$. On the other hand, the $\sim$38.21\degree~structure belongs to the symmetry group $D^1_3$ (space group 149, P312), with its K-point belonging to the point group $C_3$. The different symmetry groups influence the splitting of the energy bands at the K-points, encoded by irreducible representations (irreps). In the twisted homobilayers, this effect is clearly visible in the spin splittings directly at the K-point, which are non-zero for all the studied bands of the $\sim$38.21\degree~case but are absent in some of the bands of the $\sim$21.79\degree~case. Nonetheless, both v$_\text{1-4}$ and c$_\text{1-4}$ present strong signatures of the radial Rashba spin texture.

\begin{figure}[H]
    \includegraphics[width=\textwidth]{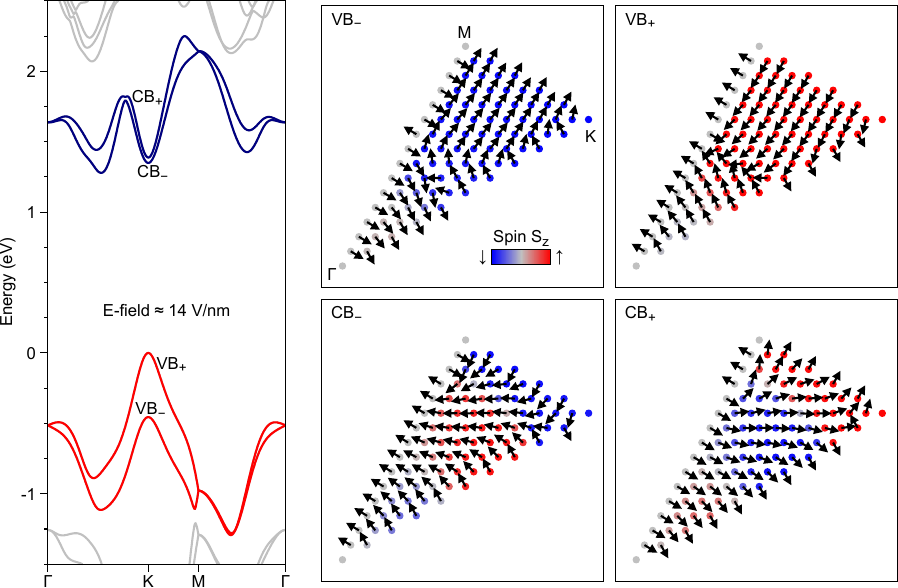}
    \caption{Monolayer WSe$_\text{2}$ under external electric field. The in-plane spin texture is normalized, i. e., $\langle\sigma_{x,y}\rangle/\sqrt{\langle\sigma_{x}\rangle^{2}+\langle\sigma_{y}\rangle^{2}}$. Values close to $\Gamma$-point are $\sim 0.5$, while values close to the $K$-point are on the order of 10$^{-2}$ - 10$^{-1}$.}
\label{fig:monoWSe2}
\end{figure}

To better understand the origin of trigonal effects in the spin texture of the twisted homobilayers as emerging from the hidden spin splitting in the individual layers, we present in Fig.~\ref{fig:monoWSe2} the band structure and spin texture for the band edges of monolayer WSe$_2$ under electric field. The electric field mimics the effect of asymmetric potentials, which is present in the twisted homobilayer system. Our calculations reveal that the in-plane spin texture exhibits complicated patterns that do not behave as typical tangential Rashba fields, particularly when we move away from the K-point. Since in the twisted homobilayers there is band hybridization involved states at arbitrary k-points, these different regions in k-space with non-tangential Rashba textures can contribute to the deviations from pure radial Rashba spin texture [as shown in Figs.~\ref{fig:twistedWSe2_21.8}(a,b)].

\cleardoublepage 

\section{Twisted Multilayer Stacks}

\begin{figure*}[htb!]
	\includegraphics[width=.99\textwidth]{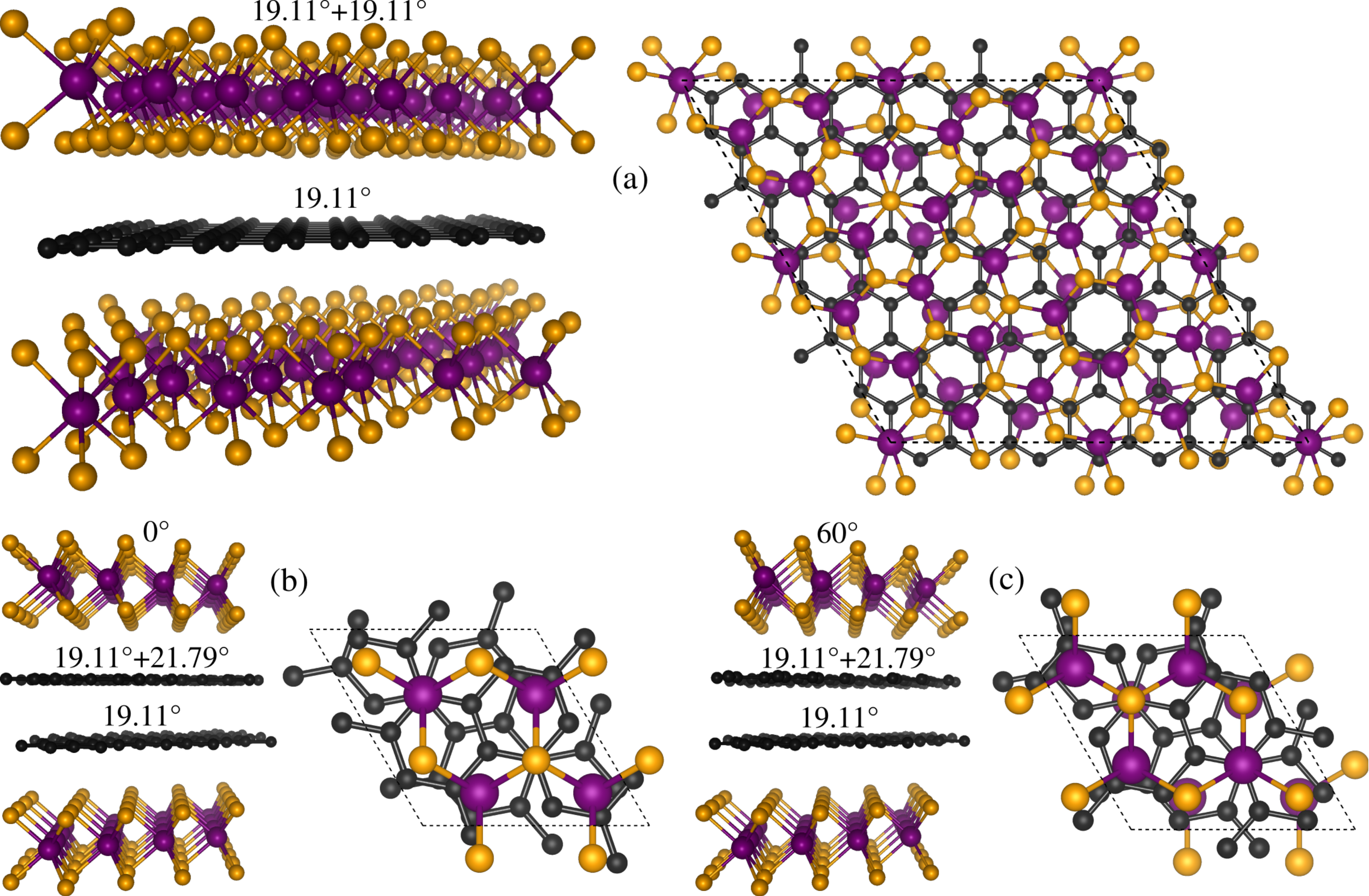}
	\caption{Top and side views of the twisted multilayer stacks. We consider (a) WSe$_2$-encapsulated graphene and (b,c) WSe$_2$-encapsulated even TBLG. 
 The twist angle indicated above each layer is measured with respect to the bottom WSe$_2$ layer. }\label{fig:multilayer_stacks}
\end{figure*}

We finally consider twisted multilayer stacks. 
In particular, we consider 
the even case of TBLG and encapsulate it within WSe$_2$ monolayers, which induce strong proximity-induced SOC. 
Two different geometries, which have 52 atoms in the heterostructure supercell, are set-up with the {\tt atomic simulation environment (ASE)} \cite{ASE} and the {\tt CellMatch} code \cite{Lazic2015:CPC}, implementing the coincidence lattice method \cite{Koda2016:JPCC,Carr2020:NRM}. In the heterostructures, the graphene layers are biaxially strained by about 0.8\%, while keeping WSe$_2$  unstrained. The structures and the twist angles are shown in Fig.~\ref{fig:multilayer_stacks}(b,c).

The \textit{ab initio} calculations are performed with {\tt Quantum ESPRESSO}~\cite{Giannozzi2009:JPCM}.
Self-consistent calculations are carried out with a $k$-point sampling of $18\times 18 \times 1$. We use an energy cutoff for charge density of $560$~Ry and
the kinetic energy cutoff for wavefunctions is $70$~Ry for the fully relativistic pseudopotentials
with the projector augmented wave method~\cite{Kresse1999:PRB} with the 
Perdew-Burke-Ernzerhof exchange correlation functional~\cite{Perdew1996:PRL}.
Spin-orbit coupling (SOC) is included in the calculations.
For the relaxation of the heterostructures, we add DFT-D2 vdW corrections~\cite{Grimme2006:JCC,Grimme2010:JCP,Barone2009:JCC} and use 
quasi-Newton algorithm based on trust radius procedure. 
Dipole corrections \cite{Bengtsson1999:PRB} are also included to get correct band offsets and internal electric fields.
To get proper interlayer distances and to capture possible moir\'{e} reconstructions, we allow all atoms to move freely within the heterostructure geometry during relaxation. Relaxation is performed until every component of each force is reduced below $1\times10^{-4}$~[Ry/$a_0$], where $a_0$ is the Bohr radius.

Additionally, we consider the 38.21\degree~twisted WSe$_2$ homobilayer, and place a monolayer graphene in between the WSe$_2$ layers. In particular, we place graphene such, that the relative twist angles from top and bottom WSe$_2$ with respect to graphene are exactly opposite, i.~e., $\pm \frac{1}{2}\times 38.21\degree$. We end up with 266 atoms in the heterostructure supercell. The calculation details are the same as before, but with a reduced $k$-grid of $3\times 3\times 1$. The structure and the twist angles are shown in Fig.~\ref{fig:multilayer_stacks}(a).

\begin{figure*}[htb]
	\includegraphics[width=.99\textwidth]{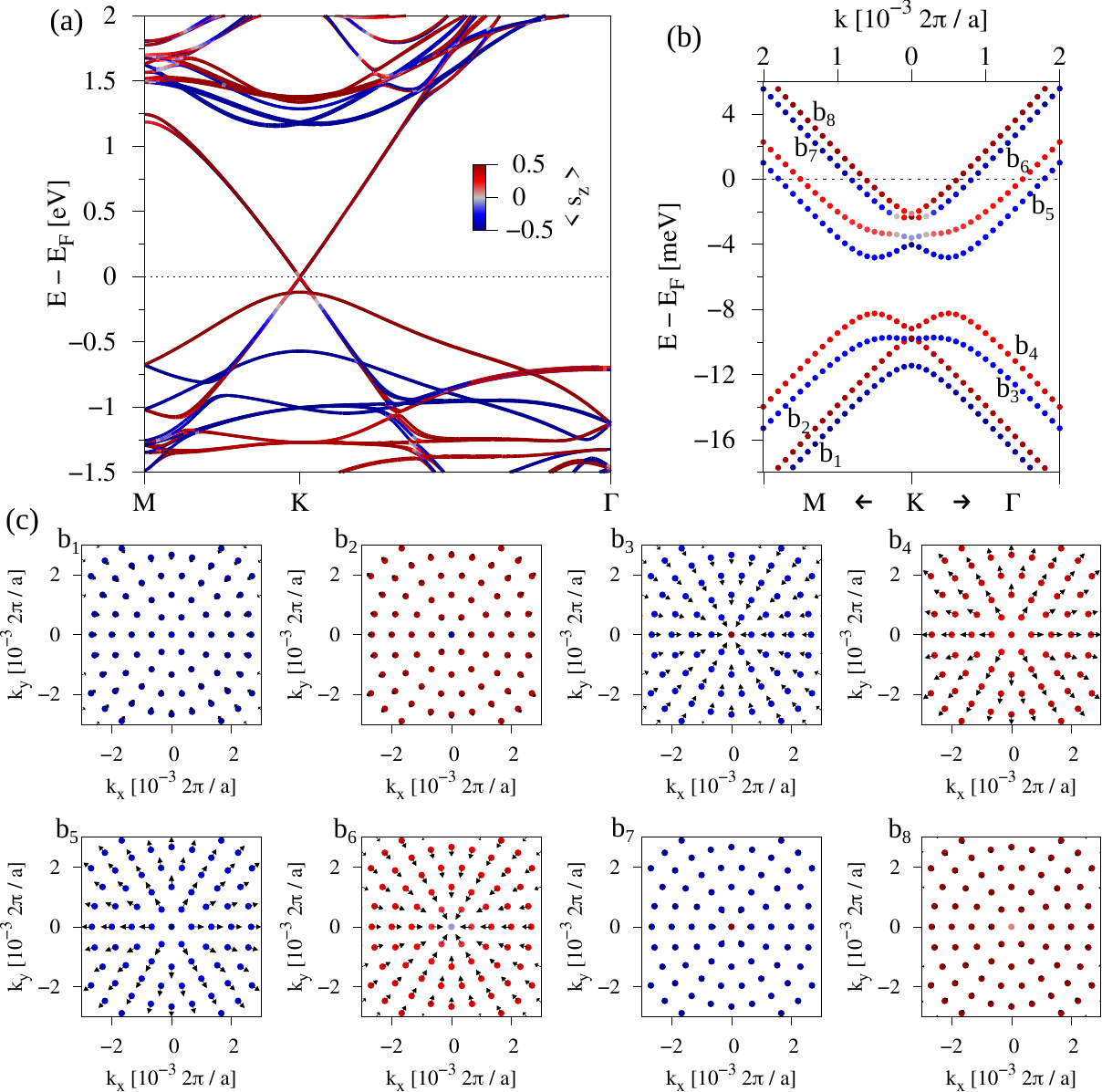}
	\caption{(a) Global band structure of the WSe$_2$-encapsulated even TBLG geometry of Fig.~\ref{fig:multilayer_stacks}(b). (b) The corresponding low energy bands with labels for the eight bands b$_1$-b$_8$. (c) The spin-orbit fields of the eight bands as indicated in (b).}
 \label{fig:TBLG_encap_conf1}
\end{figure*}

\begin{figure*}[htb]
	\includegraphics[width=.99\textwidth]{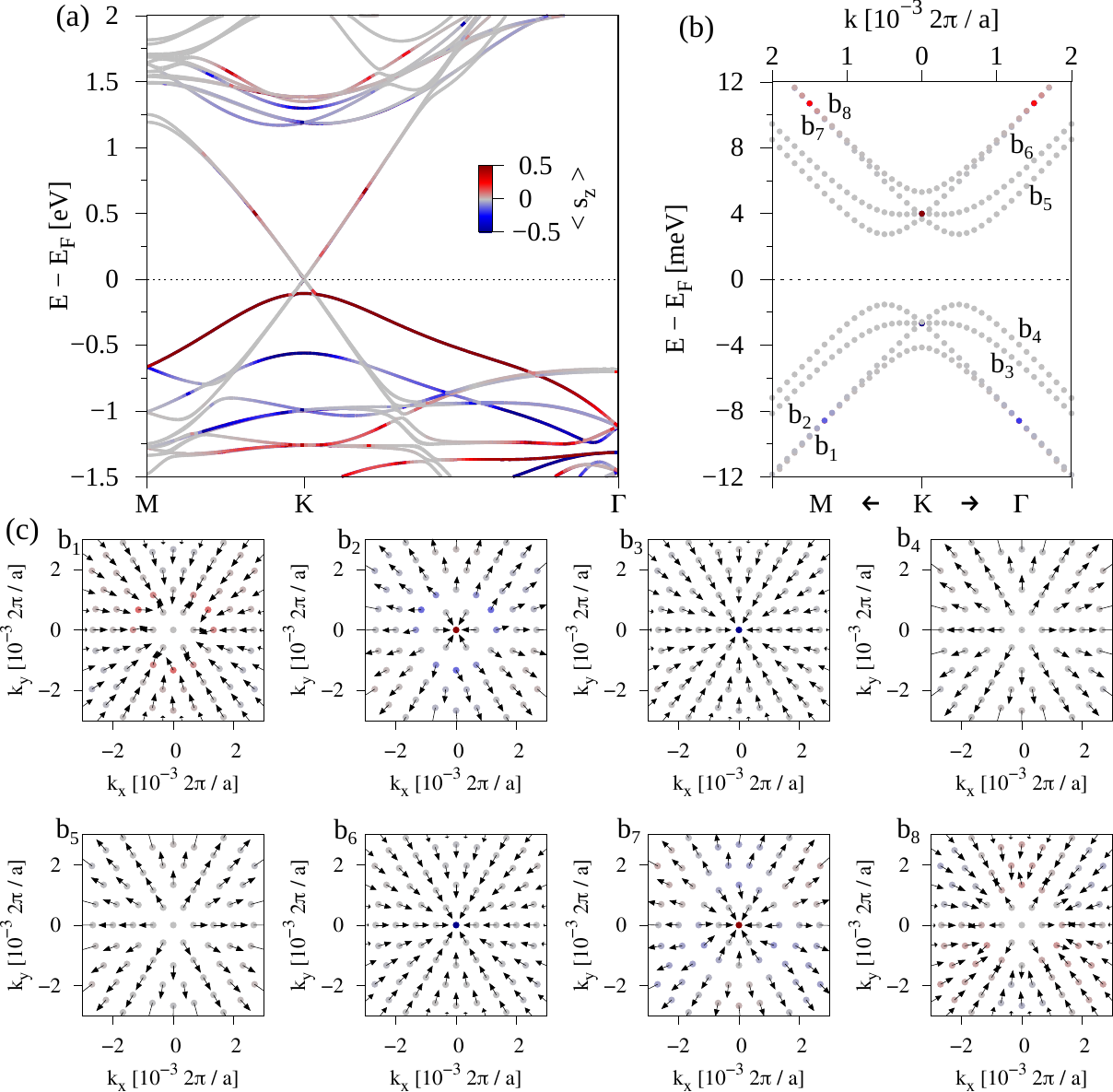}
	\caption{(a) Global band structure of the WSe$_2$-encapsulated even TBLG geometry of Fig.~\ref{fig:multilayer_stacks}(c). (b) The corresponding low energy bands with labels for the eight bands b$_1$-b$_8$. (c) The spin-orbit fields of the eight bands as indicated in (b).}
 \label{fig:TBLG_encap_conf2}
\end{figure*}

\begin{figure*}[htb]
	\includegraphics[width=.99\textwidth]{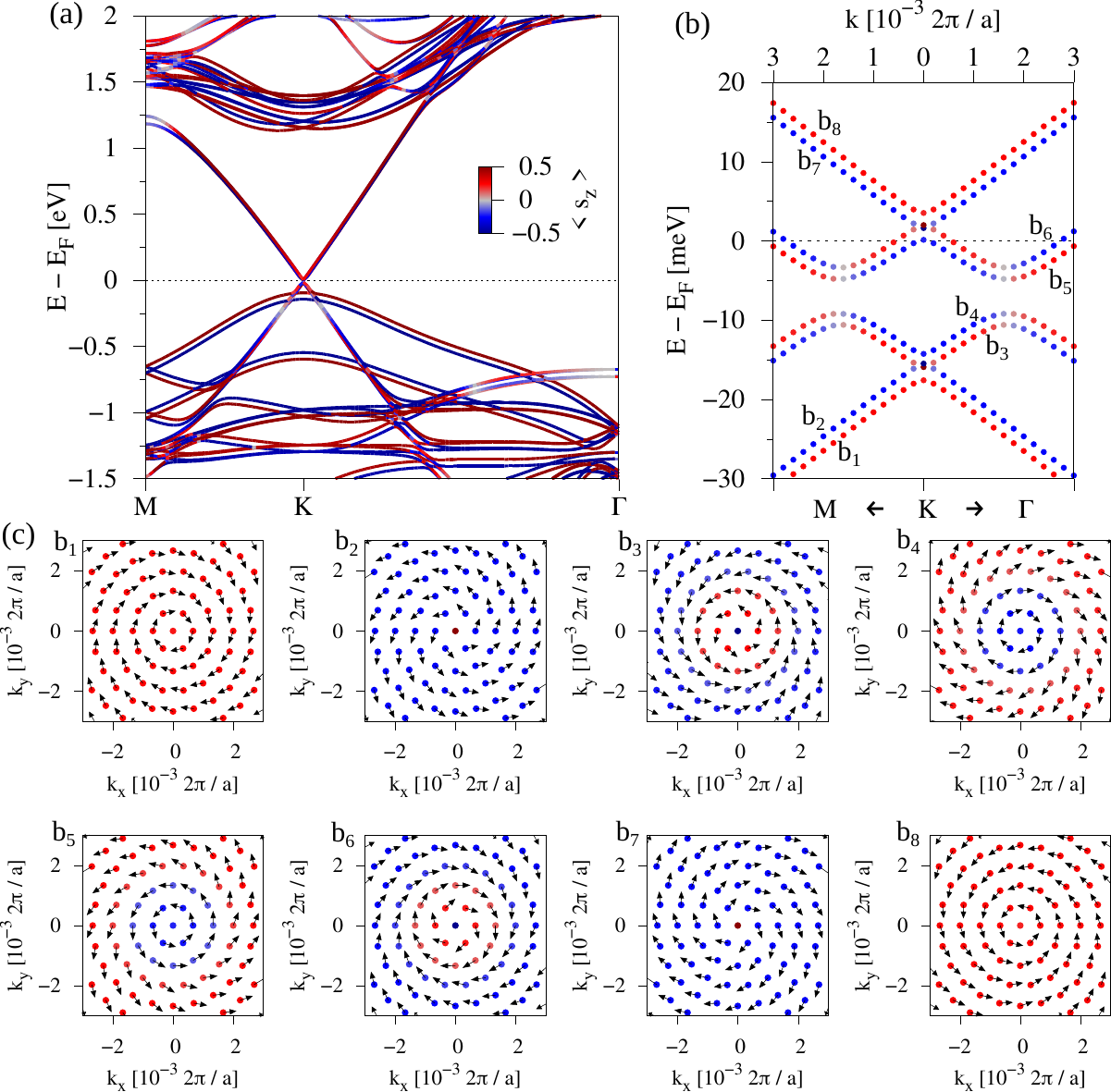}
	\caption{(a) Global band structure of the WSe$_2$-encapsulated even TBLG geometry of Fig.~\ref{fig:multilayer_stacks}(c), with an applied transverse electric field of 0.1~V/nm. (b) The corresponding low energy bands with labels for the eight bands b$_1$-b$_8$. (c) The spin-orbit fields of the eight bands as indicated in (b).}
 \label{fig:TBLG_encap_conf2_efield}
\end{figure*}

In Fig.~\ref{fig:TBLG_encap_conf1} and Fig.~\ref{fig:TBLG_encap_conf2}, we show the DFT-calculated global band structures, low energy bands, and spin-orbit fields for the WSe$_2$-encapsulated TBLG geometries corresponding to Fig.~\ref{fig:multilayer_stacks}(b,c). 
In the case when the two WSe$_2$ layers have a relative twist angle of 0\degree, out-of-plane spin components dominate the spectrum, with small but radial in-plane spin components. 
In the case of 60\degree~relative twist angle, the radial in-plane spins dominate the low energy spectrum of TBLG. By applying a small electric field of 0.1~V/nm, see Fig.~\ref{fig:TBLG_encap_conf2_efield}, we introduce a potential difference between the twisted graphene layers, and spin-orbit fields recover their typical vortex-like Rashba structure. 

\begin{figure*}[htb]
	\includegraphics[width=.99\textwidth]{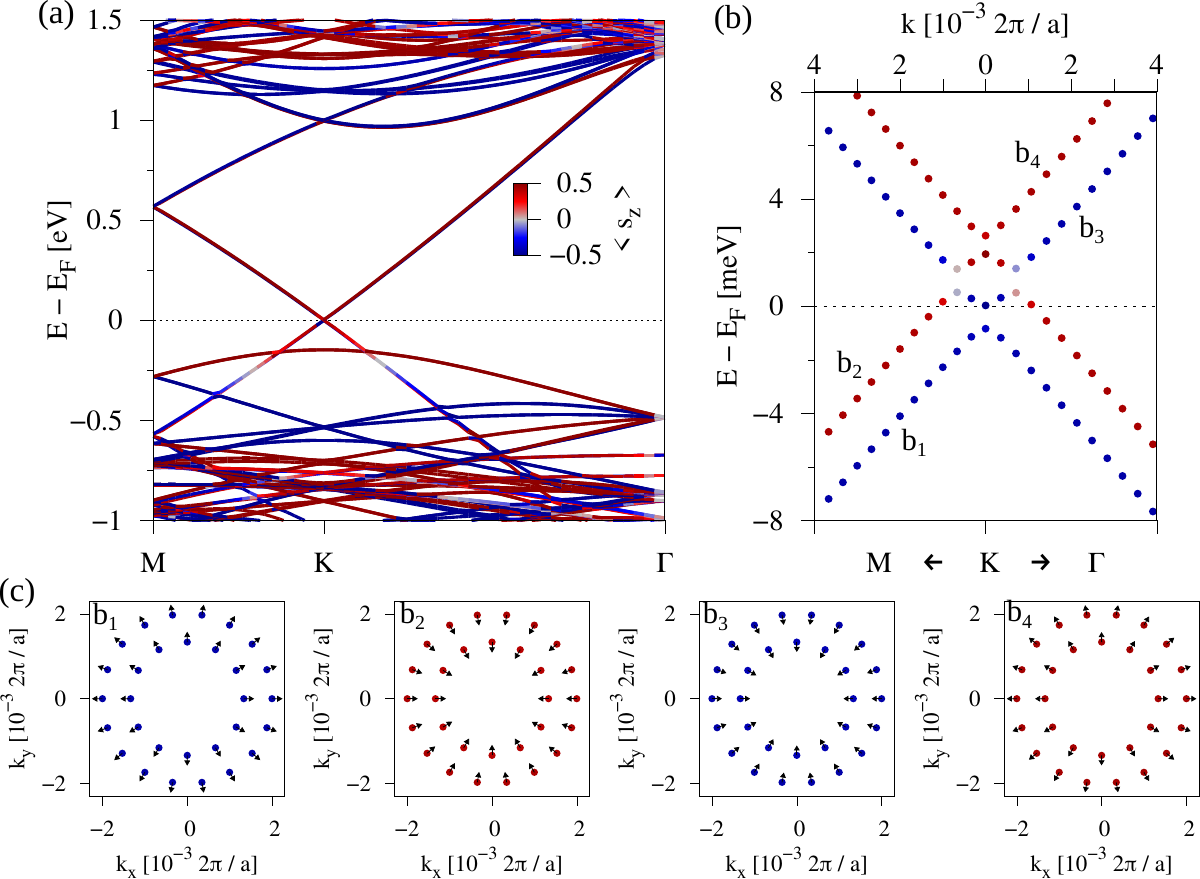}
	\caption{(a) Global band structure of the WSe$_2$-encapsulated monolayer graphene geometry of Fig.~\ref{fig:multilayer_stacks}(a). (b) The corresponding low energy bands with labels for the eight bands b$_1$-b$_4$. (c) The spin-orbit fields of the four bands as indicated in (b).}
 \label{fig:MLG_WSe2_encap}
\end{figure*}

In the case of WSe$_2$-encapsulated monolayer graphene, see Fig.~\ref{fig:MLG_WSe2_encap}, we also find a radial Rashba spin-orbit field combined with strong out-of-plane valley-Zeeman spins.

\begin{figure*}[htb]
	\includegraphics[width=.99\textwidth]{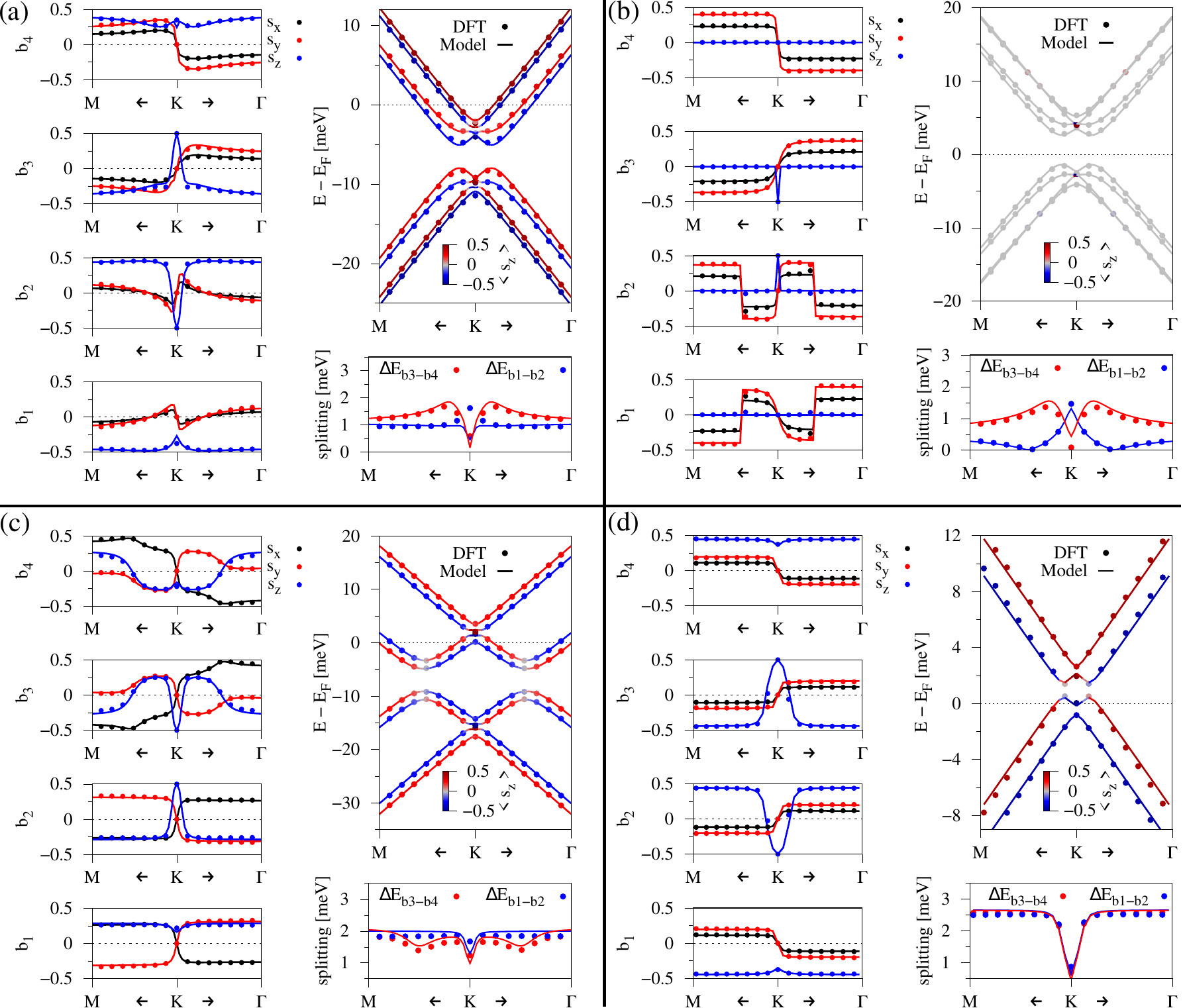}
	\caption{Comparison of the DFT data and model Hamiltonian fits. (a) Corresponds to the low energy bands in Fig.~\ref{fig:TBLG_encap_conf1}. (b) Corresponds to the low energy bands in Fig.~\ref{fig:TBLG_encap_conf2}. (c) Corresponds to the low energy bands in Fig.~\ref{fig:TBLG_encap_conf2_efield}. (d) Corresponds to the low energy bands in Fig.~\ref{fig:MLG_WSe2_encap}. }
 \label{fig:Fitting_tBLG_WSe2}
\end{figure*}

The low energy bands and spins of the encapsulated TBLG geometries, with and without electric field, can be reproduced by the model Hamiltonians from Sec.~\ref{sec:models}, employing sublattice-resolved intrinsic and complex-valued Rashba SOCs for the individual graphene layers to account for the proximity effects~\cite{Naimer2021, Lee2022:PRB,Zollner2023:PRB}.
The comparison of DFT and model data are shown in Fig.~\ref{fig:Fitting_tBLG_WSe2}, while the parameters are summarized in Table~\ref{Tab:Fit_Results_WSe2_encap_TBLG}. In the case of encapsulated graphene, we employ the following model Hamiltonian
\begin{flalign}
\label{Eq:Hamiltonian}
&\mathcal{H} = \mathcal{H}_{0}+\mathcal{H}_{\Delta}+\mathcal{H}_{\textrm{I}}+\mathcal{H}_{\textrm{R}}+E_D,\\
&\mathcal{H}_{0} = \hbar v_{\textrm{F}}(\kappa k_x \sigma_x - k_y \sigma_y)\otimes s_0, \\
&\mathcal{H}_{\Delta} =\Delta \sigma_z \otimes s_0,\\
&\mathcal{H}_{\textrm{I}} = \kappa (\lambda_{\textrm{I}}^\textrm{A} \sigma_{+}+\lambda_{\textrm{I}}^\textrm{B} \sigma_{-})\otimes s_z,\\
&\mathcal{H}_{\textrm{R}} = -\lambda_{\textrm{R}}\textrm{e}^{-\textrm{i}\theta_{\textrm{R}}\frac{s_z}{2}}(\kappa \sigma_x \otimes s_y + \sigma_y \otimes s_x)\textrm{e}^{\textrm{i}\theta_{\textrm{R}}\frac{s_z}{2}}.
\end{flalign}

\begin{table*}[htb]
\caption{\label{Tab:Fit_Results_WSe2_encap_TBLG} The fit parameters of the model Hamiltonians. The model parameters from left to right reproduce the DFT data in the subfigures (a)-(d) of Fig.~\ref{fig:Fitting_tBLG_WSe2}.
Compared to the Hamiltonians from the main text, we employ layer (1, 2) and sublattice (A, B) resolved intrinsic and Rashba SOC parameters to account for proximity effects. 
}
\begin{ruledtabular}
\begin{tabular}{c c c c c  }
system & TBLG &  TBLG &  TBLG &  graphene\\
relative twist angle of WSe$_2$ layers [deg.] & 0 & 60 & 60 & 38.21 \\
E-field [V/nm] & - & - & 0.1 & -\\
 \hline
$\Delta$~[meV] & - & - & - & 0.015 \\
$u$ [meV] & 0 & 0 & 8.329 & -\\
$w$ [meV] & 3.486 & 3.604 & 3.604 & - \\ 
$v_{\textrm{F}}/10^5 [\frac{\textrm{m}}{\textrm{s}}]$ & 7.940 & 7.931 & 7.931 & 7.456 \\
$\phi$ [rad] & -0.654 & -0.721 & -0.721 & - \\
$\lambda_{\textrm{I}}^\textrm{A1}$~[meV] & 0.567 & 0.567 & 0.567 & 1.133 \\ 
$\lambda_{\textrm{I}}^\textrm{B1}$~[meV] & -0.615 & -0.615 & -0.615 &  -1.229 \\
$\lambda_{\textrm{I}}^\textrm{A2}$~[meV] & 0.567 & -0.615 & -0.615 & - \\ 
$\lambda_{\textrm{I}}^\textrm{B2}$~[meV] &-0.615 & 0.567 & 0.567 &  - \\
$\lambda_{\textrm{R1}}$~[meV] & 0.865 & 0.865 & 0.875 & 0.601 \\
$\lambda_{\textrm{R2}}$~[meV] & -0.865 & -0.865 & -0.855 & - \\
$\theta_{\textrm{R1}}$ [rad]  & 0.866 & 0.866 & 0.866 & -1.571 \\
$\theta_{\textrm{R2}}$ [rad] & 1.229 & 1.229 & 1.229 & - \\
$E_{\textrm{D}}$ [meV]  & -6.541 & 0.605 & -6.996 & 0.943 \\
\end{tabular}
\end{ruledtabular}
\end{table*}

\clearpage

\newpage

\bibliography{paper}